# Towards a Family-based Analysis of Applicability Conditions in Architectural Delta Models

Arne Haber[1], Thomas Kutz[1], Holger Rendel[1],
Bernhard Rumpe[1], and Ina Schaefer[2]

[1] Software Engineering, RWTH Aachen University, Germany
[2] Institute for Software Systems Engineering, TU Braunschweig, Germany

**Abstract.** Modeling variability in software architectures is a fundamental part of software product line development. $\Delta$-MontiArc allows describing architectural variability in a modular way by a designated core architecture and a set of architectural delta models modifying the core architecture to realize other architecture variants. Delta models have to satisfy a set of applicability conditions for the definedness of the architectural variants. The applicability conditions can in principle be checked by generating all possible architecture variants, which requires considering the same intermediate architectures repeatedly. In order to reuse previously computed architecture variants, we propose a family-based analysis of the applicability conditions using the concept of inverse deltas.

## 1 Introduction

Modeling variability of the software architecture is an integral part in software product line development. $\Delta$-MontiArc [11] is a modular, transformational variability modeling approach for software architectures. In $\Delta$-MontiArc, a family of software architectures is described by a designated core architecture model and a set of delta models containing modifications to the core architecture. A delta model can add and remove components, ports and connections and modify the internal structure of components. By applying the modifications contained in a delta model, an existing architecture model is transformed into another architectural variant. A particular variant in the architecture family is specified by a product configuration comprising the deltas that have to be applied to the core architecture. In order to resolve conflicts between delta models modifying the same architectural elements, an application order constraint can be attached to each delta model determining which other delta models have to be or should not be applied before this delta model.

Application order constraints are also used to ensure that each delta model is applicable to the core or intermediate architecture during product generation. Applicability means that all elements removed or modified by the delta exist and that all elements added by the delta do not yet exist. If these applicability conditions hold, the architecture resulting from delta application is defined, otherwise the result is undefined, following [17]. In order to check that the application order constraints guarantee the applicability of the delta models during product generation, a naive *product-based approach* is to generate and check the architectures for all possible product configurations and all possible intermediate architectures. This naive approach is very inefficient because for examining all possible product architectures, the same intermediate products might have to be re-generated several times.

In this paper, we propose *inverse deltas* in order to enable an efficient family-based analysis of the applicability conditions in architectural delta models. A *family-based analysis* checks all products that can be derived by traversing the whole artifacts base of the product line only once, without generating all possible products explicitly. An inverse delta reverts the operations carried out by the original delta such that applying the delta and its inverse to an architecture retrieves the original architecture. The family-based analysis constructs the *family application order tree (FAOT)* which contains all possible delta application orders that comply to the application order constraints attached to the delta models. Using inverse deltas, it is possible to traverse the $FAOT$ without generating the same intermediate architectures several times. Instead, the tree is only traversed once in a depth-first manner. In this traversal, already computed intermediate architectures are reused by reconstructing them with the application of inverse deltas. If the analysis of the $FAOT$ passes the applicability conditions checks, it is guaranteed that for all possible product configurations, which are subsets of the set of deltas models, satisfy the applicability conditions and lead to a defined resulting architecture.

This paper is structured as follows: Section 2 briefly introduces $\Delta$-MontiArc. Product Generation is described in Section 3. The family-based analysis using inverse deltas is proposed in Section 4 and discussed in 5. Section 6 describes related approaches. Section 7 concludes with an outlook to future work.

## 2 $\Delta$-MontiArc

$\Delta$-MontiArc [11] is a modular and transformational approach for describing architectural variability and is based on the textual architecture description language (*ADL*) MontiArc [10]. MontiArc focuses on the domain of distributed information-flow architectures. An example for a MontiArc architecture is given in Listing 1.1 which represents an Anti Lock Braking System (ABS). It contains inputs for four wheelsensors which measure the current speed of the four wheels of a car, a signal for the braking command, and four outputs to control the brake actuators. The component `abs` calculates the individual braking pressures for all wheels. If a wheel is close to a blocking state indicated by the corresponding wheel sensor, it reduces the braking pressure for this wheel to maintain the stability of the vehicle.

```
1 component BrakingSystem {
2   autoconnect port;
3   port
4     in WheelSensor wheelspeed1,
5     in WheelSensor wheelspeed2,
6     in WheelSensor wheelspeed3,
7     in WheelSensor wheelspeed4,
8     in BrakeCommand brake,
9     out BrakePressure wheelpressure1,
10    out BrakePressure wheelpressure2,
11    out BrakePressure wheelpressure3,
12    out BrakePressure wheelpressure4;
13   component ABS abs;
14 }
```
**Listing 1.1.** MontiArc Model for an Anti-Lock Bracking System.

```
1 delta ElectronicStabilityControl after TractionControl {
2   modify component BrakingSystem {
3     add port in AccelerationSensor lateralaccel;
4     remove component tc;
5     add component ESC esc;
6     connect lateralaccel -> esc.accel;
7   }
8 }
```
**Listing 1.2.** Delta for Electronic Stability Control.

In $\Delta$-MontiArc, MontiArc is extended with the concept of delta modeling [4, 16, 15] to represent architectural variability. Based on a core architecture specified in MontiArc, architectural deltas are specified that add, remove or modify architecture elements using the operations add, remove and modify for ports, components and corresponding parameters. For connectors, the operations connect and disconnect are available. Further possible operations are listed in [9], but not required for the comprehension of this paper.

An example for a delta model specified with $\Delta$-MontiArc is given in Listing 1.2. The depicted ElectronicStabilityControl delta can only be applied if the TractionControl delta, which adds an input for accelleration pedal position, is executed before. This information is provided by an application order constraint in an after clause which specifies deltas which must or must not be executed before the current delta (l. 1). The BrakingSystem is modified (l. 2) by adding a new input for the lateral acceleration (l. 3) and replacing the traction control subcomponent tc with subcomponent esc (l. 4). Finally, the new input is connected to the new component (l. 5).

A product configuration for a concrete product is a set of deltas which must be applied to the core model. Listing 1.3 gives an example of a product configuration for a motorbike (l. 2) which is equipped with traction control (TC, l. 3) and an

```
1 deltaconfig StreetMotorbike {
2   TwoWheel,
3   TractionControl,
4   TwoWheelTC,
5   ElectronicStabilityControl,
6   TwoWheelESC
7 }
```

**Listing 1.3.** Product configuration of a Street Bike with TC and ESC.

electronic stability control (ESC, l. 5). To adapt TC and ESC to a motorbike, additional deltas (ll. 4, 6) are needed.

## 3  Product Generation

Product generation in $\Delta$-MontiArc is the process of generating a concrete product architecture by applying selected deltas to a given core architecture. The product generator of $\Delta$-MontiArc processes three different kinds of input models. As shown in [9], at first a product configuration is needed that determines a selection of deltas to be applied for a concrete product architecture. Second, a MontiArc architecture model for the core architecture is required and, third, $\Delta$-MontiArc delta models determine variants of the core architecture.

*Product Generation Process.* Product generation is performed in four steps. At first, MontiArc models for the core architecture are loaded, and their corresponding abstract syntax tree is stored. In the second step, a product configuration is parsed, and the delta models contained in the configuration are loaded. The *generation order* of the selected delta models is computed based on the given application order constraints. When a linear generation order is determined, delta models are applied to the core architecture in the third step of product generation. All modification operations of the deltas are applied stepwise to the core architecture. To assure the definedness of the generated product architecture, the applicability of the modification operations needs to be ensured. The following *applicability conditions* [9] are necessary for the delta operations add, remove, or modify:

- A component $c$ can only be modified, if $c$ exists.
- An architectural element $ae$ must not be added to component $c$, if $c$ already contains $ae$.
- An architectural element $ae$ must not be removed from component $c$, if $c$ does not contain $ae$.
- A port $p$ must not be removed from component $c$, if $c$ contains a connector with $p$ as its source or target.
- A subcomponent $sc$ must not be removed from component $c$, if $c$ contains a connector that has a port of $sc$ as its source or target.

The application order constraints capture dependencies between deltas to ensure the validity of the applicability conditions. If the current delta modification operation satisfies the given applicability conditions, it is applied to the core model. After all delta modifications are applied, MontiArc context conditions are checked for the generated architecture that ensure its internal consistency (see [10] for a complete list of MontiArc context conditions). In contrast, the intermediate architectures are not required to be valid MontiArc architectures.

*Checking of Applicability Conditions.* The checking of the applicability conditions is closely connected to the product generation process as the applicability of one delta operation depends on the intermediate product architecture resulting from the application of all former delta operations. When generating a concrete product architecture, the respective product configuration defines which deltas are applied, and a possible generation order can be derived and checked. However, when it should be established that all possible product configurations satisfy the applicability conditions, all possible (intermediate) architectures have to be considered. In a naive product-based analysis, all product architectures are generated and analyzed separately. Thus, the same intermediate product architectures which occur in several products during product generation are repeatedly regenerated.

## 4  Family-based Analysis of Applicability Conditions

In a family-based analysis, the core architecture model and the delta models of a product line are analyzed only once, without generating all possible product architectures by applying the respective delta models to the core architecture explicitly. Instead of repeatedly generating intermediate products, intermediate products are reused which is more efficient than a naive product-based analysis.

*Family Application Order Tree.* In order to check the applicability constraints by a family-based analysis, a *family application order tree (FAOT)* is created. The *FAOT* for the example introduced in Section 2 is shown in Figure 1. In a *FAOT*, the nodes represent the deltas of the product line. Each path in the *FAOT* starting from the root is a generation order that is valid according to the application order constraints attached to the deltas. The root node corresponds to the core architecture indicating that no delta has yet been applied and combines the forest of possible generation orders into a tree. Leaves of the *FAOT* correspond to maximal possible generation orders, where the addition of another delta will violate the application order constraints of the deltas on the path to the leaf. To each node in the *FAOT*, an architecture is associated that is generated by applying the deltas leading to this node including the node itself to the core architecture. This architecture is either a product architecture that is valid according to the MontiArc context conditions or an intermediate architecture.

The applicability conditions of the deltas can be checked by traversing all paths in the *FAOT* and establishing the applicability conditions for each modification operation encountered. In this way, all possible product architectures are analyzed that can be generated by a (sub-)path in the *FAOT*. Due to a large number of deltas and a sparse set of application order constraints, the *FAOT* can be fairly complex. Therefore, the efficient computation of the intermediate products that are necessary to traverse the *FAOT* is essential. Two approaches can be distinguished:

1. The intermediate architectures for certain tree nodes are stored such that they can be reused for calculating further intermediate products which, however, requires a hugh amount of memory for large architecture models.
2. Inverse deltas can be applied to a generated architecture to undo the application of a delta in order to backtrack in the *FAOT* without storing intermediate architectures.

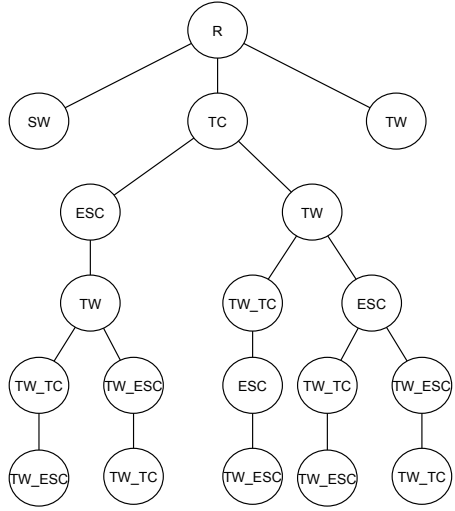

**Fig. 1.** *FAOT* for the example product line

*Inverse Deltas.* For each delta model $D$ consisting of a set of delta operations, there is an inverse delta model $D^{-1}$ such that for any product architecture $P$, it holds that $apply(apply(P, D), D^{-1}) = P$ where application of the modification operations in a delta is defined by $apply : Arch \times Delta \rightarrow Arch$ for $Arch$ the set of MontiArc architectures and $D$ the set of delta models in $\Delta$-MontiArc. An inverse delta $D^{-1}$ is derived from a delta $D$ by inverting each delta modification operation in $D$ and also inverting the ordering of the modification operations. For each `add` statement, the inverse operation is a `remove` statement and, vice versa. The operations `connect` and `disconnect` statements are inverses for each other. The enclosing component modification operations remain unchanged. An example for an inverse delta is shown in Listing 1.4.

*FAOT Analysis Using Inverse Deltas.* Using inverse deltas, the *FAOT* can be traversed in a depth-first manner without storing any intermediate architectures

```
delta A {                          | delta A_Inverse {
  modify component Base {          |   modify component Base {
    add port Integer p;            |     disconnect p -> sub.input;
    connect p -> sub.input;        |     remove port p;
  }                                |   }
}                                  | }
```

**Listing 1.4.** Inverting the delta on the left side results in the delta on the right side.

at the *FAOT* nodes. For the *FAOT* depicted in Figure 1, the following depth-first traversal is computed: $SW \to SW^{-1} \to TC \to ESC \to TW \to TW\_TC \to TW\_ESC \to TW\_ESC^{-1} \to TW\_TC^{-1} \to TW\_ESC...$

The applicability conditions are checked by processing the deltas during the traversal one by one. After processing one delta, its inverse is computed, if necessary, and stored for later application. Since an inverse delta depends on the intermediate architecture to which the original delta is applied, it is not possible to compute all required inverse deltas up front.

## 5  Discussion

Comparing the family-based analysis using inverse deltas to an analysis in which all intermediates architectures at decision-nodes in the *FAOT* are stored (intermediate approach), the inverse delta approach requires less memory. In large product lines where the *FAOTs* contain many decision nodes, memory space might become a severe problem, as every intermediate architecture that has to be stored comprises the ASTs of the modified core architecture. Compared to the naive product-based approach, the inverse delta approach uses the same amount of memory, as in both approaches no intermediate products are stored.

Regarding runtime complexity, the worst case is if there are $n$ deltas without application order constraints. Then, every possible permutation of deltas is contained in the *FAOT*. The amount of edges in a *FAOT* is $AE(n) = \sum_{i=0}^{n-1} \frac{n!}{i!} = n! * \sum_{i=0}^{n-1} \frac{1}{i!}$. The intermediate approach computes every delta once, such that $AE(n)$ steps are needed to check every possible product. The inverse delta approach visits every edge twice, once applying a delta, and once applying its inverse. The most right path of the *FAOT* is visited only once. Thus, in the inverse delta approach $2 * AE(n) - n = n! * (2 * \sum_{i=0}^{n-1} \frac{1}{i!} - \frac{1}{(n-1)!})$ steps are necessary where $\sum_{i=0}^{\infty} \frac{1}{i!} = e$ is a constant factor, and for $n \to \infty$, the term $\frac{1}{(n-1)!}$ converges to zero. Hence, both factors, as well as the constant factor of 2, may be neglected for an estimation of complexity such that the inverse delta approach as well as the intermediate approach belong to complexity class $O(n!)$ The naive product-based analysis approach generates $n!$ products by applying $n$ deltas for each product. In total, $n*n!$ delta applications are performed leading to a complexity of $O(n*n!)$. This yields that the family-based analyses are about n times faster in the worst-case than the product-based analysis. Nevertheless, a complexity of $O(n!)$ is still very high, but this is accounted to the inherent complexity of family-based analyses.

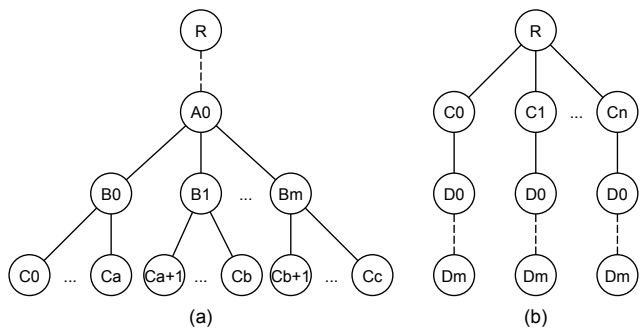

**Fig. 2.** *FAOTs* with late and early decision-nodes

The shape of the *FAOT* influences the number of inverse deltas that are required to get from one leaf to the next. Figure 2 shows two examples. Tree (a) contains many decision-nodes close to the leafs. So deriving $C1$ based on $C0$ is done in two steps by applying the inverse delta $C0^{-1}$ and afterwards delta $C1$. In contrast, tree (b) contains only one decision-node (the root). To get from the product on the very left to the next product whose path starts with $C1$, $m$ inverse deltas have to be applied. As the root node is the only decision-node, no intermediate architectures have to be stored such that the inverse delta approach is about $2*m$ steps slower without saving any memory. Accordingly, we suggest a hybrid approach that considers the shape of the *FAOT* and stores intermediate architectures at selected decision-nodes. This way, some backtracking steps with inverse deltas can be omitted such that a balance between memory consumption and runtime effort can be achieved. For example, consider the worst-case *FAOT* and assume that we store intermediate architectures at the last decision-nodes before the leaves. On level $n-2$ of the *FAOT*, each node has two children which each has one child that are leaves, since there are only 2 more deltas left which have to be applied. Storing these intermediate architectures saves 4 inverse delta applications for each of the nodes on level $n-2$, except for the most right node where only 2 steps will be saved. On level $n-2$, we have $\frac{n!}{2}$ nodes such that a reduction of $4*\frac{n!}{2} - 2 = 2n! - 2$ inverse delta applications can be achieved with only minor increase in memory consumption.

## 6   Related Approaches

Architectural variability modeling approaches can be classified into annotative, compositional and transformational modeling approaches. Annotative approaches, e.g., [6], consider one model representing all products and define which parts of the model are removed to derive a product model. Compositional approaches, e.g., [1], associate model fragments with product features that are composed for a particular feature configuration. Transformational approaches, such as CVL [12], represent variability of a base model by rules describing how a base model is transformed in order to obtain a particular product model. $\Delta$-MontiArc can be classified as a transformational approach.

Product line analyses can be classified in three main categories [18]: first, product-based analyses consider each product variant separately. Second, feature-based analyses consider the building blocks of the different product variants in isolation to derive results about all variants, but in general rely on heavy restrictions on the admissible product line variability. Third, family-based analyses check the complete code base of the product line in a single analysis to obtain a result about all possible variants.

Family-based product line analyses are currently used for type checking [2, 7] and model checking [5, 8, 14] of product lines. The approach presented in [2] also constructs all possible application orders of feature modules (which are comparable to delta models in our approach) and checks that in any possible combinations of feature modules all required references are provided. The type checking approach proposed in [7] uses a constraint-based type system where a large formula is constructed from the product line's feature model and the feature module constraints that is true if all product variants are type safe. In [17], the type safety of all product variants is checked based on the analysis of a product abstraction that is generated from constraints derived for delta modules. Thus, it can be classified as an mixture between product- and feature-based analyses.

Storing only the differences between products, as we do with inverse deltas, is also applied in versioning systems. For instance, the Revision Control System (RCS) [19] only keeps the most recent version and a sequence of inverse modifications in order to retrieve prior versions which is more efficient than working with complete version snapshots. The formalization of DARCS patch theory [13] has a concept of inverses although on a fairly abstract level. In recent work, Batory et al. [3] apply the idea of differencing for updating a program obtained by feature-oriented composition. However, in that approach it is unclear whether the differences can be expressed by means of feature modules, while in the approach presented in this paper, inverse deltas can be expressed by the same linguistic means as ordinary deltas.

## 7 Conclusion

The family-based analysis to validate $\Delta$-MontiArc product lines is an extension of our previous work [11, 9]. In this paper, we have introduced the concept of inverse deltas that allows traversing the FAOT without storing intermediate architectures. For future work, we are planning to evaluate the proposed approach at large case examples. Furthermore, we will extend the inverse delta approach to deal with the convenience operations presented [9], such as the replacement of components, where there is no obvious inverse.